# Resolving Complex Subwavelength Grating Structures Using Topologically Structured Light

Thomas A. Grant, Eric Plum, Nikolay I. Zheludev, and Kevin F. MacDonald*

Optoelectronics Research Centre, University of Southampton,
Highfield, Southampton, SO17 1BJ, UK

* Correspondence to: kfm@orc.soton.ac.uk

**It has been seen recently that when probing a nanoscale object to determine, for example, size or position via light scattering, significant advantage in measurement precision can be gained from exploiting phase singularities in a topologically structured incident light field. Here, we demonstrate that this advantage, derived from the dependence of scattered intensity profiles on strong local (subwavelength-scale) intensity and phase variations in the incident field, can be extended towards imaging applications: Analysis of scattering patterns from arbitrary binary gratings under superoscillatory illumination successfully resolves feature sizes down to ~$\lambda/7$ in single-shot measurements (a factor of 1.4× smaller than is achieved with plane wave illumination), and ~$\lambda/10.5$ in positionally-displaced multi-shot measurements (which yields no improvement in the plane wave case). Interestingly, there are circumstances in which more complex objects are better resolved than simple structures, because interference effects increase the information content of their scattering patterns.**

Over recent decades, the spatial resolution achievable in far-field optical imaging has advanced well beyond the classical Abbe diffraction limit of ~ $\lambda/2$ (where $\lambda$ is the wavelength of light), through the use of various deterministic, stochastic, and computational signal processing techniques [1-12]. In positional and dimensional optical metrology, techniques based on interferometric free-space light and evanescent field scattering or beam deflection have been developed for tracking isolated (typically optically trapped) nanoparticles with sub-nanometric precision [13-18], and a range of approaches to linear translation measurement with similar precision, leveraging light fields structured at sub-wavelength scales (by metasurfaces, plasmonic nanostructures, and spatial light modulators), have been reported [19-22]. Indeed, precision and accuracy reaching the atomic scale (100-200 pm; < $\lambda/5000$) have recently been demonstrated in single-shot optical measurements based on the deep learning analyses of objects' diffraction patterns under illumination by topologically structured light [23-26]. This is made possible by: (a) constraining the problem – i.e. the parameter space on which the inverse scattering problem is solved (using a neural network [27]) is reduced to the retrieval of one, or not more than a few, dimensional parameters from simple, well-defined objects (e.g. the widths and separation of a pair of nano-rods [26], or the position of a nanowire [25]); (b) the fact that the information content of a nano-object's diffraction pattern can be orders of magnitude larger when it is illuminated with a topologically structured light field containing phase singularities (i.e. high phase and intensity variations with deeply subwavelength scales), as opposed to a plane wave [28].

Here, we examine the extent to which the advantage of topologically structured illumination can be retained in the more challenging task of retrieving dimensional parameters from arbitrarily structured objects, i.e. as the dimensionality of the parameter space grows and feature sizes decrease. From single-shot diffraction patterns of random, barcode-like, one-dimensional gratings under superoscillatory illumination, a trained neural network consistently retrieves features down to a size of $\sim\lambda/7$ – a factor of 1.4× smaller than is achieved with plane wave illumination. This advantage is extended to a factor of 2.2× in few-shot superoscillatory imaging (with shot-to-shot object translation), whereby feature sizes down to $\sim\lambda/10.5$ are resolved. Somewhat counterintuitively, it is found that retrieval success rates can be higher for more complex objects, because interference between light scattered by multiple (as compared to few) object features increases the information content of the scattered field.

Measuring the position or displacement of a nanowire as reported in Refs. [24, 25] is a task that can be identically described as retrieving the dimensions of a double slit: the widths $A$ and $B$ of slits on either side of the nanowire, under the constraint that $A + B$ is constant. As target objects for the present computational study, we consider random grating patterns of an arbitrary number of transparent slits in an otherwise perfectly opaque, absorbing screen, illuminated with monochromatic light at a wavelength $\lambda = 488$ nm (Fig. 1). We generate a set of 4096 different grating profiles, each 10 μm long with 10 nm pixelation, as follows: In each case, the first pixel is set as either transparent or opaque with equal probability; Subsequent pixels then either adopt the same state or change to the opposing state with a probability randomly sampled from a flat distribution of values from 0.01 to 0.11; After any change from transparent to opaque or vice versa, a new change probability is selected. This ensures that feature sizes (meaning the distance between any two changes of state) within a single pattern, and across the whole set of patterns, are uncorrelated but span a range of values of interest around and below the classical diffraction limit (from $\sim\lambda/2$ down to $\lambda/50$).

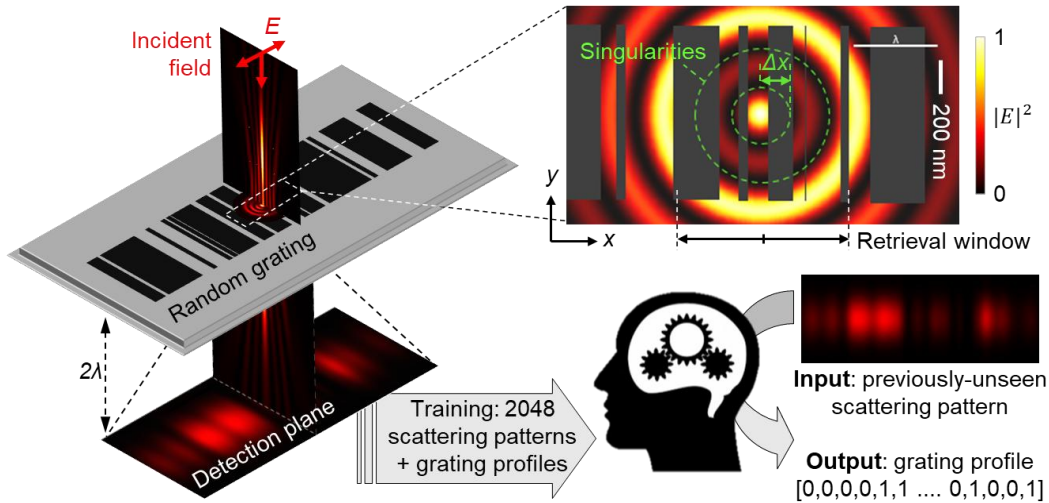

**Figure 1: Resolving arbitrary subwavelength grating structures using topologically structured light.** Arbitrary – randomly generated – binary gratings are illuminated with a superoscillatory light field, the intensity profile of which is illustrated top right (overlaid with a representative grating pattern). Dimensional profiles of the central (1 μm wide) section of gratings are retrieved via a deep learning-enabled analysis of their transmission scattering patterns.

We assume the gratings are illuminated with an axially symmetric superoscillatory light field [24-26, 29, 30] formed by the linear combination of two band-limited, prolate spheroidal wave functions (PSWFs): $\widetilde{U}(r/\lambda) = [21.65 S_2(r/\lambda) + S_3(r/\lambda)]$, where $r$ is the distance from the beam axis. This gives an incident field profile with a central intensity peak of $0.3\lambda$ half-maximum width, flanked by a series of phase singularities. Taking this to be centered on the gratings, their transmission scattering patterns at a distance of $2\lambda$ beyond the sample are calculated using the angular spectrum method, assuming a 5.12 µm × 1.28 µm field of view with 10 nm pixelation. From these scattering patterns we seek to retrieve the dimensional profile of the central 1 µm-long section of the grating, in the form of a 1×100 vector of zeros (=opaque) and ones (=transparent). To this end, half of the set of 4096 scattering patterns and corresponding grating profiles, selected at random, are used for training and validation of a convolutional neural network. This network utilizes two convolutional layers [4×4 and 2×2 kernel sizes], both of which are followed by a pooling layer [3×3 and 2×2 kernel sizes], followed by two fully-connected layers with output sizes 128 and 100, respectively. The remaining 2048 scattering patterns are then used for testing, i.e. as unseen patterns for nominally unknown grating profiles. For comparison, the exercise is repeated for plane wave (i.e. unstructured) illumination, using the same set of gratings split into the same training/validation and testing subsets. In both cases, results are aggregated over ten independent neural network training and testing cycles.

Figures 2a-f shows an illustrative selection of retrieved vs. actual grating profiles, including: relatively simple and sparse profiles (a, b) – i.e. containing a small number of well-separated feature 'edges' (transparent/opaque transitions); profiles containing few but closely-spaced edges (c, d) – i.e. very narrow features, down to $\sim\lambda/50$ wide; and more complex profiles containing a higher numbers of features with a variety of dimensions (e, f).

For direct imaging methods, the Rayleigh criterion is used to define resolution [31]. However, it is not appropriate for a probabilistic multi-parameter retrieval problem, since retrieving a feature of a certain size in one case does not imply that features of that size will be retrieved in all cases (e.g. compare Figs. 2c, d). As such, we quantify resolving power in terms of the mean rate at which pixels are correctly retrieved, as a function of the size of the feature within which they are located (Fig. 2g). Retrieval is all but perfect for feature sizes above the classical $\lambda/2$ diffraction limit, with both plane wave and superoscillatory illumination. Below this, smaller features can be resolved in both cases because interference between light scattered by different object features translates information about short length-scale structure into that part of the angular spectrum which propagates into the far-field, and the neural network training process provides a deconvolution mechanism to access such information. A retrieval success rate >95% is maintained with plane wave illumination down to feature sizes of 100 nm ($\sim\lambda/4.9$), and with superoscillatory illumination down to 70 nm ($\lambda/7$).

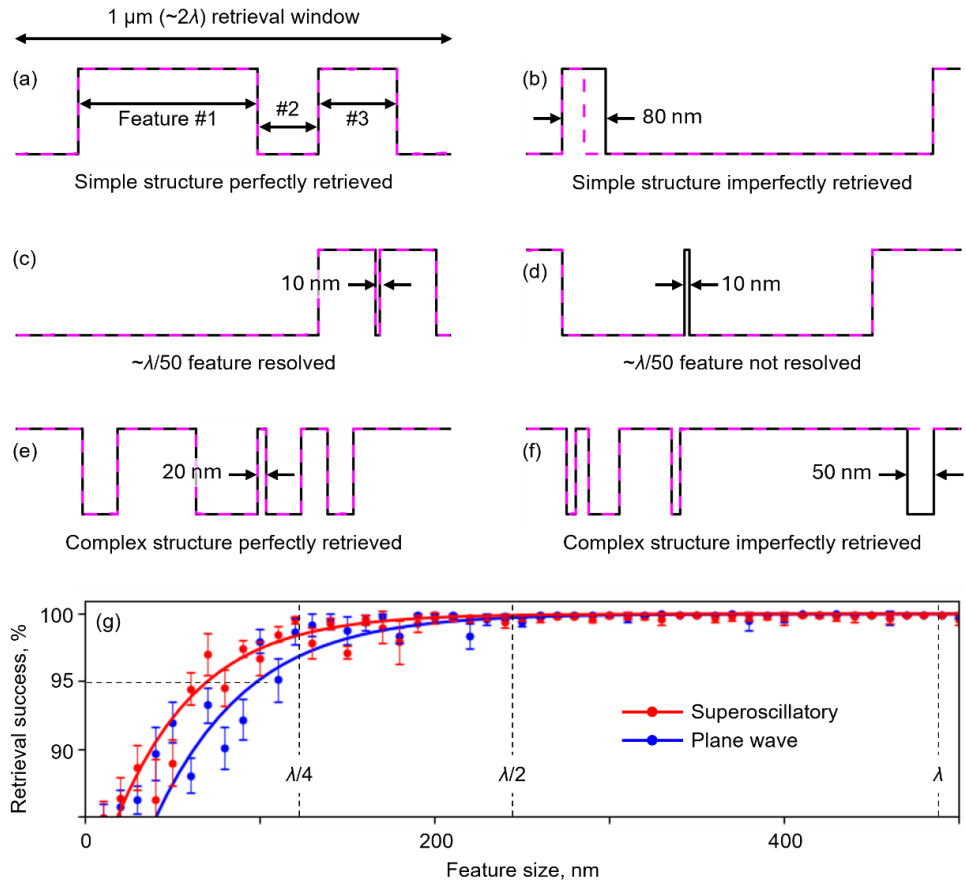

**Figure 2: Resolving arbitrary grating features beyond the diffraction limit.** (a-f) Illustrative examples of perfect and imperfect grating profile retrieval with superoscillatory illumination, for profiles of varying complexity, containing features of varying size. Solid black lines are the true grating profiles, pink dashed lines are optically retrieved profiles. g) Retrieval success as a function of feature size for plane wave and superoscillatory illumination. Points and error bars represent respectively the mean and standard deviation of results from ten independently trained networks.

The enhancement of resolution in the superoscillatory case, by a factor of ~1.4, is explained by the greater information content [32, 33] of the more complex scattered field, arising from the presence of high local (subwavelength scale) intensity and phase variations in the incident field interacting with the object, particularly in the vicinity of phase singularities [28]. The increased complexity of the scattering pattern does however also make the retrieval problem more challenging, and this is reflected in neural network learning rates: With superoscillatory illumination, the network takes more epochs to converge but does so to a lower mean squared error, thereby ultimately yielding higher retrieval success rates for smaller feature sizes.

Greater superoscillatory advantage can be gained from the fact that diffraction patterns of topologically structured light are highly sensitive to changes in the mutual positions of object features (e.g. grating slit edges) and incident field features (esp. phase singularities). Indeed, it has been demonstrated that this sensitivity can provide for optical localization of a known single nano-object with picometric (i.e. atomic scale) precision at visible wavelengths [24, 25]. Thus, for the purposes of parameter retrieval from arbitrary objects, scanning or multi-shot recording of diffraction patterns (i.e. with known shot-to-shot translation between the incident superoscillatory beam and the object) will provide additional information for neural network training and subsequent retrieval of grating profiles from unseen scattering patterns. Figure 3 shows how retrieval performance is enhanced through the use of just two additional diffraction patterns, recorded for incident beam displacements of $\pm \Delta x$ (where $\Delta x$ is equal to the distance between the beam axis and the first singularity in the radial profile, $\sim \lambda/3$ – see Fig. 1). In this case, the 95% threshold for successfully resolving grating features extends down to $\sim \lambda/10.5$. No such enhancement of performance is possible under plane wave illumination because scattering patterns in that case are invariant with respect to translation of the incident field.

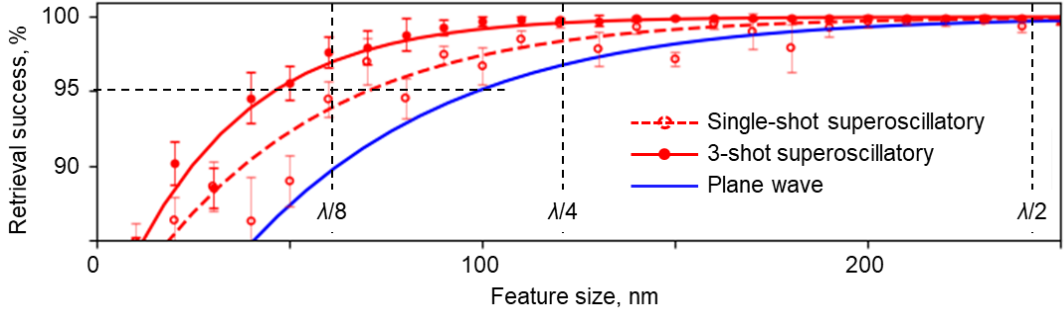

**Figure 3: Multi-shot enhancement of resolution.** Retrieval success as a function of feature size for single- and 3-shot superoscillatory illumination of gratings. In the single-shot case, the incident beam is centred at (0,0); In the 3-shot case, additional diffraction patterns are recorded at beam positions $\pm \Delta x \sim \lambda/3$. Points and error bars represent respectively the mean and standard deviation of results from ten independently trained networks. (The trend line for plane wave illumination from Fig. 2 is overlaid for reference).

Thus far, we have considered retrieval performance only as a function of feature size, regardless to the complexity of the grating, i.e. the number of features within the 1 μm retrieval window. Interestingly, single-shot retrieval performance is found to be better – meaning smaller features are retrieved with greater success – for more complex objects (Fig. 4). Here, we define 'simple' gratings as those presenting between one and three features of any size (i.e. 2-4 opaque/transparent transitions – see Fig. 2a) within the retrieval window. 40% of the 2048 set of used for testing are in this category. Another 40%, i.e. those presenting four or more features, are classified as 'complex' (only 6% present ≥8 features; none present >12 features). The remaining 20% contain only a single edge (or are empty) and as such do not include what we define as a 'feature' in the present case. We attribute this performance differential, which is greater for plane wave illumination (than for superoscillatory illumination), to the fact that information is carried from the object plane to the detection plane through interference, whereby the total amount of interference, and hence the total information content of the scattered field increases with the number of scattering features [34]. Of course, the associated improvement in retrieval success seen in Fig. 4 cannot continue indefinitely: with finite detector resolution and finite dataset size, and where increasing complexity equates to decreasing feature size (within a finite retrieval window), there must come a point where the deconvolution mechanism fails. Indeed, as shown in Fig. 1g, retrieval success falls rapidly (regardless of grating complexity) for feature sizes $<\lambda/10$.

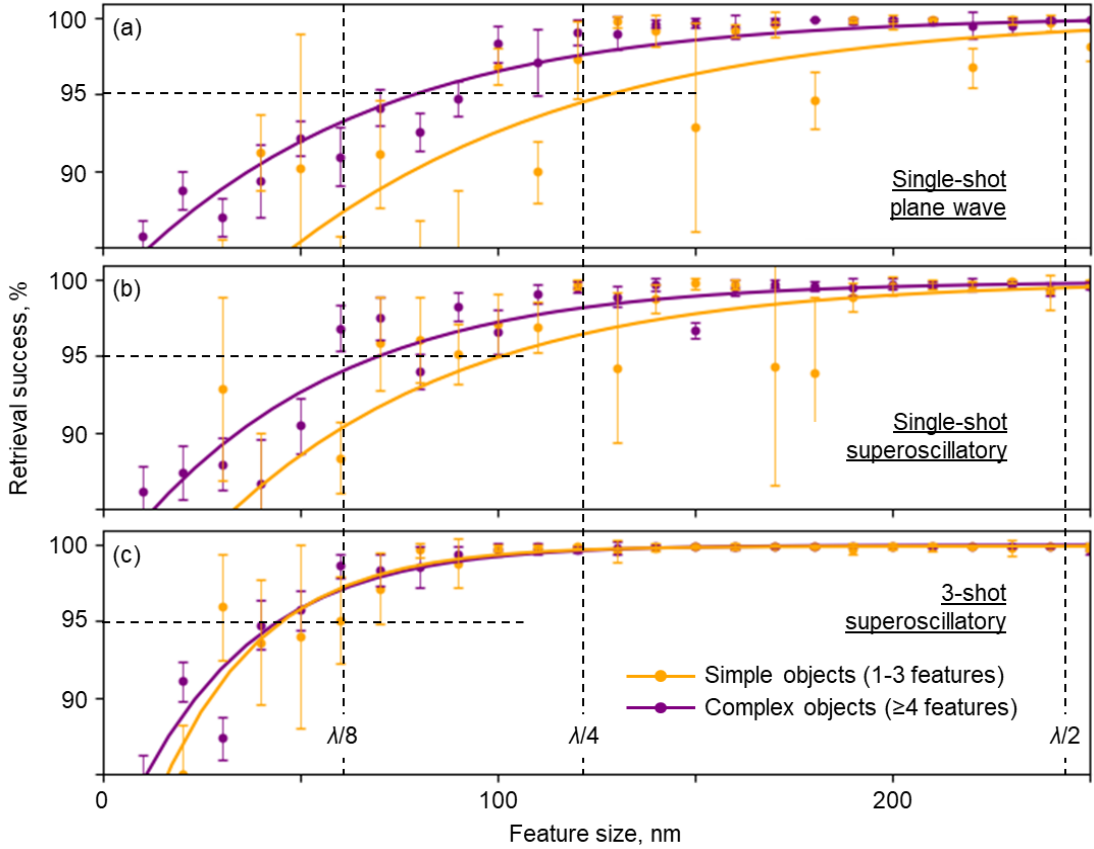

**Figure 4: Parameter retrieval from simple vs. complex objects.** Retrieval success as a function of feature size for simple gratings [presenting 1-3 features within the 1 μm retrieval window – orange points and trend lines] and complex gratings [≥4 features – purple points and trend lines] for (a) single-shot plane wave, (b) single-shot superoscillatory, (c) 3-shot superoscillatory illumination. Points and error bars represent respectively the mean and standard deviation of results from ten independently trained networks.

In summary, we have shown here that the advantages of topologically structured illumination (over plane wave illumination), previously seen in scattering-based optical metrology, are maintained when the dimensional parameter space expands from that of a tightly-constrained single value retrieval task to that of retrieving an unspecified number of parameters from a complex object. Despite the increased complexity of the parameter estimation problem, neural networks can establish a deconvolution algorithm which leverages an enhancement of information content in scattering patterns derived from the interaction of strong local intensity and phase variations in the incident field with nanometric features of the scattering object. Single-shot resolution down to $\lambda/7$ is achieved in retrieval of random binary grating profiles under superoscillatory illumination, improving to $\lambda/10.5$ in few-shot imaging

(beating plane wave illumination to a factor of 2.2×). These results open a path to the use of topologically structured light for increasingly complex nanoscale optical metrology and basic (e.g. binary object) imaging tasks.

**Acknowledgements**

This work was supported by the UK Engineering and Physical Sciences Research Council (grant EP/T02643X/1).

**Data availability**

All information necessary to reproduce the results presented is contained within the article.